\documentclass[twocolumn]{svjour3}

\journalname{Journal of Mathematical Biology}

\usepackage{amssymb,amsfonts}
\usepackage{amsmath,mathtools}
\usepackage{graphicx}
\usepackage{color}
\usepackage{bm}
\usepackage{float}
\usepackage{algorithm}
\usepackage{algpseudocode}

\RequirePackage{fix-cm}
\smartqed

\newcommand{\heading}[1]{\smallskip\noindent\textbf{\textit{#1}}}

\usepackage[
backend=biber,
style=numeric-comp,
sorting=none
]{biblatex}
\begin{document}
\title{Learning biological neuronal networks with artificial neural networks: neural oscillations}
\author{Ruilin Zhang$^{1,2,*}$ \and Zhongyi Wang$^{1,3,*}$ \and Tianyi Wu$^{1,3}$ \and Yuhang Cai$^{4}$ \and Louis Tao$^{1,5,\dagger}$ \and Zhuo-Cheng Xiao$^{6,\dagger}$ \and Yao Li$^{7,\dagger}$}

\institute{  $^1$ Center for Bioinformatics, National Laboratory of Protein Engineering and Plant Genetic Engineering, School of Life Sciences, Peking University, Beijing, 100871, China; \\
             $^2$ Yuanpei College, Peking University, Beijing, 100871, China; \\
             $^3$ School of Mathematical Sciences, Peking University, Beijing 100871, China; \\
             $^4$ Department of Mathematics, University of California, Berkeley, CA, USA 94720; \\
             $^5$ Center for Quantitative Biology, Peking University, Beijing, 100871, China; \\
             $^6$ Courant Institute of Mathematical Sciences, New York University, NY, USA 10003; \\
             $^7$ Department of Mathematics and Statistics, University of Massachusetts Amherst, MA, 01003. \\
             $^*$ Equal contribution to this paper. \\
             $^\dagger$ Corresponding authors. \\ 
              L.~Tao \email{taolt@mail.cbi.pku.edu.cn}; \\
              Z.-C.~Xiao \email{xiao.zc@nyu.edu}; \\
              Y.~Li \email{yaoli@math.umass.edu}
}

\date{Received: date / Accepted: date / Edited: date}

\maketitle

\begin{abstract}
First-principles-based modelings have been extremely successful in providing crucial insights and predictions for complex biological functions and phenomena. However, they can be hard to build and expensive to simulate for complex living systems. On the other hand, modern data-driven methods thrive at modeling many types of high-dimensional and noisy data. Still, the training and interpretation of these data-driven models remain challenging. Here, we combine the two types of methods to model stochastic neuronal network oscillations. Specifically, we develop a class of first-principles-based artificial neural networks to provide faithful surrogates to the high-dimensional, nonlinear oscillatory dynamics produced by neural circuits in the brain. Furthermore, when the training data set is enlarged within a range of parameter choices, the artificial neural networks become generalizable to these parameters, covering cases in distinctly different dynamical regimes. In all, our work opens a new avenue for modeling complex neuronal network dynamics with artificial neural networks.
\end{abstract}


\section{Introduction}
The last few decades have seen rapid developments of first-principles-based mathematical models to study living systems. Based on a collection of \textit{a priori} physiological and physical principles, the evolution of mathematical models can offer significant advantages in understanding, reproducing, and predicting complex biological phenomena.
However, first-principle-based models can be prohibitively expensive to build due to the large number of parameters and variables characterizing the complexity of biological details, e.g.,  multiple time scales, complicated interactions between biological elements, among others. 
Alternatively, modern data-driven models focusing on phenomenological or empirical observations are gaining ground in mathematical biology, in that they are designed to deal with high dimensional and noisy data \cite{janes2006data,hasenauer2015data,solle2017between,jack2018data,alquraishi2021differentiable}. However, one is still faced with the daunting task of making sense of the coordinates and parameters of the data-driven models to identify interpretable and biologically meaningful features. 

In this study, we investigate how the combination of the two classes of methods can be used to study spiking neuronal networks (SNNs). 
SNNs are capable of producing highly nonlinear, high-dimensional, and multi-timescale dynamics, and have been widely used to investigate cortical functions and their computation principles (see, e.g., \cite{ghosh2009spiking,ponulak2011introduction,nobukawa2017chaotic,borgers2003synchronization,chariker2016orientation}). First-principles-based model reduction methods such as coarse-graining and mean-field theories have been developed to better understand SNN dynamics \cite{wilson1972excitatory,brunel1999fast,buice2007field,cai2006kinetic,cai2021model,li2019stochastic}.
On the other hand, artificial neural networks (ANN, and its offspring, deep neural networks, or DNN) are a modern data-driven method inspired by the nervous system. DNN has been extremely successful in both engineering applications (image processing, natural language processing, etc.) and applied mathematics (parameter estimation, numerical ordinary/partial differential equations, inverse problem, etc.). See \cite{aggarwal2018neural,schmidhuber2015deep,chon1997linear,li2020nett,raissi2019physics}, for instance. In particular, it has been shown recently that DNN can well approximate both finite and infinite-dimensional operators \cite{barron1994approximation,kovachki2021universal}. The idea of using DNN surrogates in models to replace the firing rate of SNN was first explored by \cite{zhang2020dnn}. This motivates us to propose \textbf{a first-principle-based deep-learning framework} that replaces high dimensional mappings within neuronal network dynamics by artificial neurons. 

The neuroscience problem we address in this paper is the $\gamma$-band oscillations, a type of 30-90 Hz oscillatory physiological dynamics prominent in many brain regions \cite{HenrieShapley2005,BroschEtAl2002,bauer2006tactile,BuschmanMiller2007,MedendorpEtAl2007,van2010learning,CsicsvariEtAl2003,PopescuEtAl2009,vanderMeerRedish2009}. Remarkably, in previous studies, $\gamma$-oscillations can be produced in  simple, idealized SNN models involving only two neural populations, excitatory (E) and inhibitory (I) \cite{chariker2018rhythm,zhang2014coarse,rangan2013emergent,li2019well}. More specifically, due to transient noise and/or external stimulus, highly correlated spiking patterns (previously termed multiple-firing events, or MFEs) are repeatedly produced from the competitions between E/I populations, involving the interplay between multiple timescales.
MFEs are a type of stochastic, high-dimensional emergent phenomena, with rapid and transient dynamical features that are sensitive to the biophysical parameters of the network. Therefore, it is a very challenging task to build model reductions that can provide biological insights for $\gamma$-oscillations in a wide range of parameter regimes.


This paper explores learning the complex $\gamma$-oscillations with first-principle-based DNNs. 
Our previous study revealed that the complex $\gamma$-oscillatory dynamics can be captured by a Poincare mapping $F$ projecting the network state at one initiation of an MFE to the next initiation \cite{cai2021model}. 
Therefore, $F$ is a high-dimensional mapping subjected to biophysical parameters of SNNs, and thus very hard to analyze.
Instead, we approximate $F$ by \textbf{A.} using coarse-graining (CG) and discrete cosine transform (DCT) to reduce the dimensionality of the state space, and \textbf{B.} benefiting from the representation power of DNNs. Specifically, DNNs provide a unified framework for varying SNN model parameters, revealing the potential of generalization to different dynamical regimes of the emergent network oscillations.
Despite the significant underlying noise and the drastic dimensional reductions, our DNNs successfully capture the main feature of the $\gamma$-oscillations in SNNs.
This effectively makes the DNN a surrogate of the true biological neuronal networks. 


The organization of this paper is as follows. Section 2 introduces the neuronal network model that serves as the ground truth. The descriptions and capturing algorithm of MFEs are depicted in Section 3. Section 4 discusses how to set up the training set for artificial networks. The main results are demonstrated in Section 5. Section 6 is the conclusion and discussion.

\section{Neuronal network model description}
\label{Sect2: MIFModel}

Throughout this manuscript, we study SNN dynamics with a Markovian integrate-and-fire (MIF) neuronal network model. This model imitates a small, local circuit of the brain and shares many features of local circuits in living brains, including extensive recurrent interactions between neurons, leading to the emergence of $\gamma$-oscillations. We will evaluate the performance of DNNs based on their predictive power of dynamics produced by the MIF model.

\subsection{An Markovian spiking neuronal network}
We consider a simple spiking network consisting of $N_E$ excitatory (E) neurons and $N_I$ inhibitory (I) neurons, homogeneously connected. The membrane potential ($V$) of each neuron is governed by Markovian dynamics, with the following assumptions:

\begin{enumerate}
  \item $V$ takes value in a finite, discrete state space;
  \item For neuron $i$, $V_i$ is driven by both external and recurrent E/I inputs from other neurons, through the effect of the arrival of spikes (or action potentials);
  \item A spike is released from neuron $i$ when $V_i$ is driven to the firing threshold. Immediately after that, neuron $i$ enters the refractory state before resetting to the rest state;
  \item For a spike released by neuron $i$, a set of {\it post-synaptic} neurons is chosen randomly. Their membrane potentials are driven by this spike.
\end{enumerate}
We now explain these assumptions in detail. 

\heading{Single neuron dynamics.} Let us index the $N_E$ excitatory neurons from $1,\cdots,N_E$, and the $N_I$ inhibitory neurons from $N_E+1,\cdots,N_E+N_I$. For neuron $i$ ($i=1,...,N_E+N_I$), the membrane potential $V_i$ lies in a discrete state space $\Gamma$
$$
V_i\in\Gamma:=\{-M_r, -M_r+1,\cdots,-1,0,1,\cdots, M\}\cup\{\mathcal{R}\},
$$
where the states $M_r$, $M$ and $\mathcal{R}$ are the inhibitory reversal potential, the spiking threshold, and the refractory state, respectively. Once a neuron is driven to the threshold $M$, its membrane potential $V_i$ enters the refractory state $\mathcal{R}$. After an exponentially distributed waiting time with mean $\tau_\mathcal{R}$, $V_i$ is reset to the rest state $0$. 

Within the state space $\Gamma$, $V_i$ is driven by the external and recurrent inputs to neuron $i$. Yet, while a neuron is in the refractory state $\mathcal{R}$, it does not respond to any stimuli. The external (i.e., from outside the network itself) stimulus serves as an analog of feedforward sensory input, e.g., from the thalamus or from other brain regions. In this paper, the external inputs to individual neurons are modeled as series of impulsive kicks, whose arrival times are drawn from independent \& identical Poisson processes. 
The rates of the Poisson processes, $\lambda^{E,I}$, are taken to be constants across the E/I populations.
Each kick received by neuron $i$ increases $V_i$ by $1$.

An E/I neuron will {\it spike} when its membrane potential $V_i$ reaches threshold $M$, sending an E/I kick to its postsynaptic neurons (the choice of which will be discussed momentarily). Each recurrent E spike received by neuron $i$ takes effect on $V_i$ after an independent, exponentially distributed time delay $\mathbf{\tau}\sim{\rm Exp}(\tau^{E})$. The excitatory spikes received by neuron $i$ that have not yet taken effect form \textit{a pending E-spike pool}, with size $H_i^{E}$. 
Therefore, $H_i^{E}$ increases by 1 when an an E kick arrives at neuron $i$, and drops by 1 when a pending spike takes effect. This discussion applies to the $I$ spikes as well: the size of pending I-spike pool is $H_i^I$, and waiting time of the pending spikes are subjected to $\mathbf{\tau}\sim{\rm Exp}(\tau^{I})$. 
In summary, the state of neuron $i$ is therefore described by a triplet $$(V_i, H_i^E, H_i^I).$$
We note that the pool sizes $H_i^E$ and $H_i^I$ may be viewed as E and I synaptic currents of neuron $i$ in the classical leaky integrate-and-fire neuron model \cite{gerstner2014neuronal}.


\heading{Impacts of spikes. }
The effects of the recurrent E/I spikes on membrane potentials are different. When a pending E-spike takes effect, $V_i$ is increased by $[S^{Q,E}]+u^E$, where 
\begin{align*}
    u^E\sim{\rm Bernoulli}(p)\quad \textrm{and} \quad p&=S^{Q,E}-[S^{Q,E}],
\end{align*} 
where $Q\in\{E,I\}$ and $[\cdot]$ denotes the floor integer function. 
Likewise, when a pending I-spike takes effect,  $V_i$ is decreased by $[S^{Q,I}]+u^I$, where 
\begin{align*}
    u^I\sim{\rm Bernoulli}(q)\quad \textrm{and} \quad q&=S^{Q,I}-[S^{Q,I}],
\end{align*} 
$V_i$ is strictly bounded the state space $\Gamma$. Should $V_i$ exceed $M$ after an E-spike increment, it will reset to $\mathcal{R}$ and neuron $i$ spikes immediately. On the other hand, should $V_i$ go below $-M_r$ due to an I-spike, it will stay at $-M_r$ instead.


\heading{A homogeneous network architecture.}
We close our discussion of the MIF model by clarifying the choice of network architecture. Instead of having a predetermined network architecture with fixed synaptic coupling strengths, the postsynaptic neurons of each spike are decided \textit{on-the-fly}. That is to say, a new set of postsynaptic neurons is chosen independently for each spike. More specifically, when a type-$Q'$ neuron spikes, the targeted postsynaptic neurons in the $Q$ populations, excluding the spiking neuron itself, are chosen with probabilities $P^{QQ'}$ ($Q,Q'\in\{E,I\}$).
We point out that the motivation of this simplification is for analytical and computational convenience by making neurons interchangeable within each subtype, and is standard in many previous theoretical studies \cite{cai2006kinetic,brunel1999fast,wilson1972excitatory,buice2007field,cai2021model,gerstner2014neuronal}.

\vspace{0.3cm}
To summarize, the state space of the network is denoted as $\mathbf{\Omega}$. A {\it network state } $\omega \in \mathbf{\Omega}$ consists of $3(N_E+N_I)$ components
\begin{align}
\nonumber
\omega=(&V_1,\cdots,V_{N_E},V_{N_E+1},\cdots,V_{N_E+N_I},\\ 
\nonumber
        &H^E_1,\cdots,H^E_{N_E},H^E_{N_E+1},\cdots,H^E_{N_E+N_I},\\
\label{Eq1-networkState}
        &H^I_1,\cdots,H^I_{N_E},H^I_{N_E+1},\cdots,H^I_{N_E+N_I}). 
\end{align}
\subsection{Parameters used for simulations}
The choices of parameters are adopted from our previous studies \cite{li2019stochastic,cai2021model,wu2022multi}. For all SNN parameters used in the simulations, we list their definitions and values in Table \ref{Table1:Parameters}. Here, we remark that the projection probability between neurons $P^{Q'Q}$ ($Q,Q'\in\{E,I\}$) are chosen to match the anatomical data in the macaque visual cortex, see \cite{chariker2016orientation} for reference. Also, $\tau^E<\tau^I$, since it is known that the Glu-AMPA receptors act faster than the GABA-GABA receptors, with both on a time scale of milliseconds \cite{koch1999biophysics}.

On the other hand, four parameters concerning recurrent synaptic coupling strength will be tested and varied in this study. This is because they are sensitive for SNN dynamics, and yet hard to directly measure by current experiments methods \cite{chariker2016orientation,xiao2021data}.
The range of tested parameter set $\mathbf{\Theta}=\{S^{EE},S^{EI},S^{IE},S^{II}\}\subset\mathbb{R}^4$ is also given by Table \ref{Table1:Parameters}.

\begin{table*}[htbp]
\begin{center}
    \begin{tabular}{|l|c|l|l|}
      \hline
      Parameter Group  & Parameter        & Meaning                       & Value \\ \hline
      Network          & $N^{E}$          & Number of E cells             & 300   \\
      architecture     & $N^{I}$          & Number of I cells             & 100   \\
                       & $P^{EE}$         & E-to-E coupling probability   & 0.15  \\ 
                       & $P^{EI}$         & I-to-E coupling probability   & 0.50  \\ 
                       & $P^{IE}$         & E-to-I coupling probability   & 0.50  \\ 
                       & $P^{II}$         & I-to-I coupling probability   & 0.40  \\
                       & $S^{EE}$         & E-to-E synaptic weight        & $[3.5, 4.5]$  \\ 
                       & $S^{EI}$         & I-to-E synaptic weight        & $[-2.5, -1.5]$  \\
                       & $S^{IE}$         & E-to-I synaptic weight        & $[2.5, 3.5]$  \\
                       & $S^{II}$         & I-to-I synaptic weight        & $[-2.5, -1.5]$  \\\hline
      Neuronal         & $M$              & Threshold potential           & 100 \\ 
      physiology       & $-M_r$           & Inhibitory reversal potential & -66 \\
                       & $\tau^\mathcal{R}$ & Expectation of refractory period & 3 ms \\
                       & $\tau^E$         & Expectation of E-spike pending time & 2 ms \\
                       & $\tau^I$         & Expectation of I-spike pending time & 4 ms \\
                       & $\lambda^E$      & Total external spikes/s to E    & 3 kHz    \\
                       & $\lambda^I$      & Total external spikes/s to I    & 3 kHz  \\ \hline
    \end{tabular}
  \label{Table1:Parameters}
 \caption{Parameters regarding the network architecture (first row) and individual neuronal physiology (second row). Symbols, meanings, and values of relevant parameters are depicted.}
\end{center}
\end{table*}



\begin{figure}[h]
    \centering
    \includegraphics[width=0.5\textwidth]{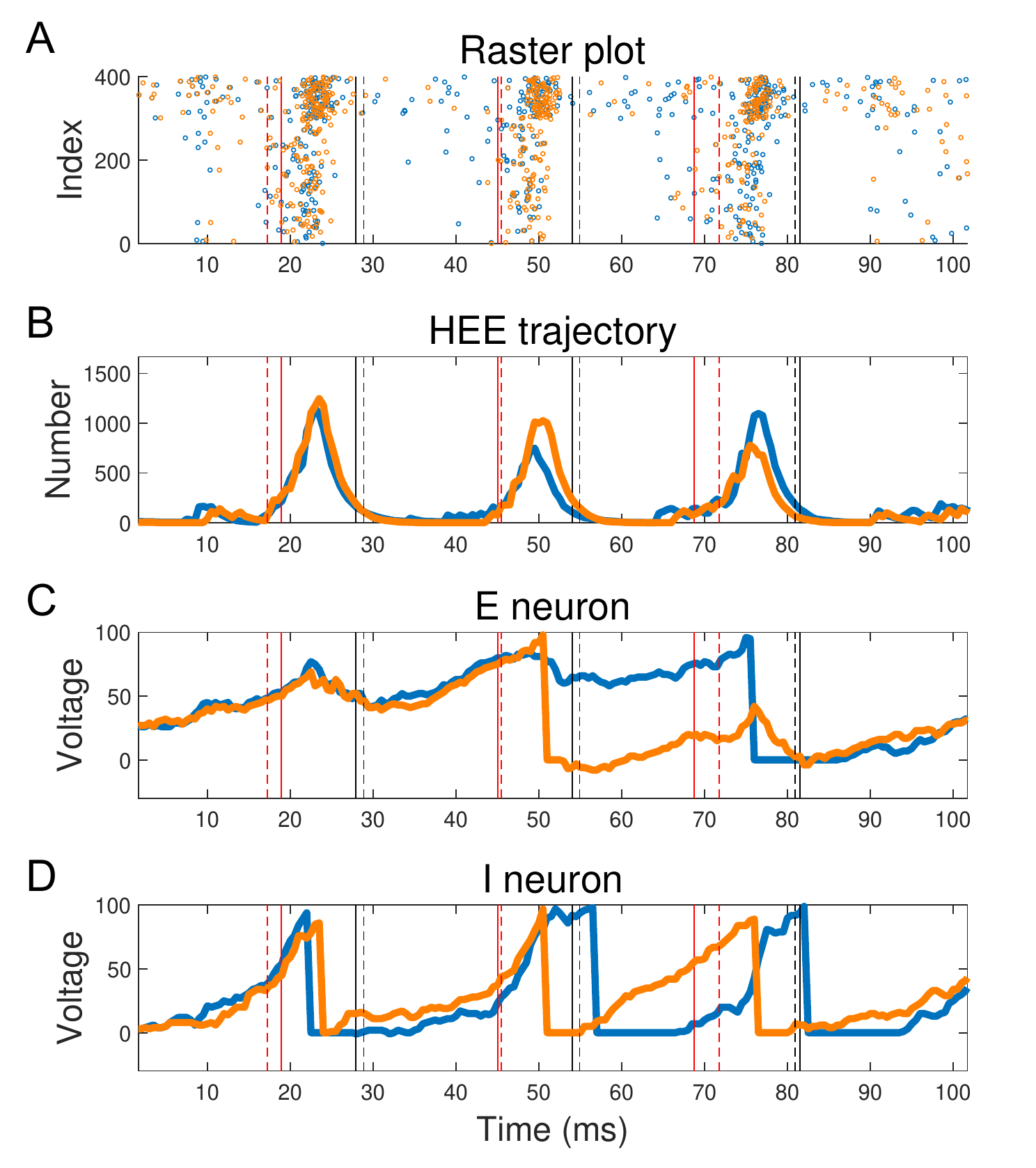}
    \caption{Divergence between SSA and tau-leaping methods. \textbf{A.} Raster plot from SSA (blue) and tau-leaping method (orange) using same seed. \textbf{B.} HEE trajectories from SSA and tau-leaping method. \textbf{C-D.} Voltage trajectories of an E/I neuron from SSA and tau-leaping method. Vertical red/black lines indicate the beginnings/endings of MFEs. (solid: SSA,  dashed: tau-leaping)}
   \label{Fig1}
\end{figure}

\section{Multiple-firing events}
\label{Sect3.1-MFEProperties}
\heading{In the studies of $\gamma$-oscillations.} 
In many brain regions, electrophysiological recordings of the local field potentials reveal temporal oscillations with power peaking in the $\gamma$-frequency band (30-90 Hz) \cite{HenrieShapley2005,BroschEtAl2002,bauer2006tactile,BuschmanMiller2007,MedendorpEtAl2007,van2010learning,CsicsvariEtAl2003,PopescuEtAl2009,vanderMeerRedish2009}. Because of the belief that these coherent rhythms play a crucial role in cognitive computations, there has been much work on understanding their mechanisms in different brain areas, in disparate dynamical regimes, and within various brain states \cite{azouz2000dynamic,AzouzGray2003,FrienEtAl2000,
WomelsdorfEtAl2012,LiuNewsome2006,FriesEtAl2001,FriesEtAl2008,bauer2007gamma,PesaranEtAl2002,WomelsdorfEtAl2007,Basar2013,krystal2017impaired,mably2018gamma}. 

To explain the neural basis of $\gamma$-oscillations, a series of theoretical models have found transient, nearly synchronous collective spiking patterns that emerge from the tight competitions between E and I populations \cite{whittington2000inhibition,rangan2013emergent,TraubEtAl2005,chariker2015emergent,chariker2018rhythm}.
More specifically, rapid \& coherent firing of neurons occurs within a short interval, and such spiking patterns recur with $\gamma$-band frequencies in a stochastic fashion. 
This phenomenon is termed multiple-firing events (MFEs). MFEs are triggered by a chain reaction initiated by recurrent excitation, and terminated by the accumulation of inhibitory synaptic currents. In this scenario, the $\gamma$-oscillations appear in electrophysiological recordings as the local change of the electric field generated by MFEs. 
We refer readers to \cite{rangan2013emergent,chariker2015emergent,chariker2018rhythm,li2019stochastic,cai2021model,wu2022multi} for further discussions of MFEs.

\heading{The alternation of fast and slow phases} is one of the most significant dynamical features of MFEs  (Fig.~\ref{Fig1}). Namely, at the beginning of an MFE, a chain reaction leads to a transient state where many neurons fire collectively in a short time interval (the fast phase). On the other hand, during the interval between two successive MFEs (an inter-MFE interval, or IMI), the neuronal network exhibits low firing rates while the system recovers from the extensive firings via a relatively quiescent period (the slow phase). These two phases can be discriminated by the temporal coherence of spikes in the raster plot where the time and location of spikes are indicated (Fig.~\ref{Fig1}A, blue dots). Here, MFEs and IMIs are separated by vertical lines - we will discuss the method of capturing MFEs momentarily in Sect.~\ref{Sect3.2-CapturingMFE}.



The complexity of MFEs is partially reflected by its sensitivity to spike-timing fluctuations \cite{xiao2022multilevel}. Specifically, during an MFE, the missing or misplacement of even a small number of spikes can significantly alter the path of network dynamics. Furthermore, the high dimensionality of state space $\mathbf{\Omega}$ and the high degree of nonlinearity make MFE hard to analyze. 
On the other hand, the network dynamics during IMIs are more robust to noise and exhibit low-dimensional features \cite{cai2021model,wu2022multi}. 

\subsection{Spike-timing sensitiveness of transient dynamics}
We illustrate this slow-fast dichotomy through a comparison between two different numerical simulations, one using the stochastic simulation algorithm (SSA) and another one using tau-leaping (Fig.~\ref{Fig1}).

\heading{Stochastic simulation algorithms vs. tau-leaping.} 
In SSA, the timings of state transitions in phase space are exact, since the next state transition is generated by the minimum of finitely many independent, exponentially distributed random variables (see Appendix).
On the other hand, algorithms simulating random processes with fixed time steps, such as tau-leaping, introduce errors to event times at every step. In a tau-leaping simulation with a fixed step size of $\Delta t$, the effect of spikes (interactions among neurons) is processed after each time step. Hence, within each time-step, all events are uncorrelated, and changes are held off until the next update.
Therefore, the precision of SSA is determined by the computational precision of the C++ code, which is much higher than tau-leaping.

\heading{Comparison.} 
Intuitively, tau-leaping methods can well capture the slow network dynamics during IMIs, but not the fast chain reactions during MFEs. The detailed comparison is depicted in Fig.~\ref{Fig1}, where we choose $\Delta t = 0.5$ ms for tau-leaping method. 
To constrain differences induced by stochastic fluctuations, we couple the two simulations with the same external noise, leaving the intrinsic noise (the random waiting times of pending spikes and refractory states) generated within the algorithms themselves.
The raster plots depict spiking events produced by both methods and diverge rapidly during MFEs (Fig.~\ref{Fig1}A. blue: SSA; orange: tau-leaping), since the crucial coherence between firing events is not accurately captured by tau-leaping. 
This point is also supported by the comparison between voltage traces of single-neurons (Fig.~\ref{Fig1}CD): Although well aligned at the beginning, the spike timing of the neurons are strongly affected by the transient fluctuation during MFEs.

In addition to the firing events, we also illustrate the comparison between pending spikes. Here we define 
\begin{align*}
    H^{Q'Q} = \sum_{i\in Q} H_i^{Q'}, \quad Q,\,Q'\in\{E,I\},
\end{align*}
i.e., the total number of pending $Q'$-spikes for type-$Q$ neurons.
Fig.~\ref{Fig1}B demonstrates the divergence of $H^{{EE}}$ collected from different simulations. Because of the identical external random noise, trajectories from two methods are similar in the first hundred milliseconds. However, very rapidly, the accumulation of errors in the first fast phase causes large divergence. Other pending spikes statistics yield similar disagreement between the two methods (data not shown).

\vspace{0.3cm}
Therefore, in the rest of this paper, we use SSA to simulate the Markov process of the SNN dynamics. 
\begin{figure*}[h]
    \includegraphics[width=\textwidth]{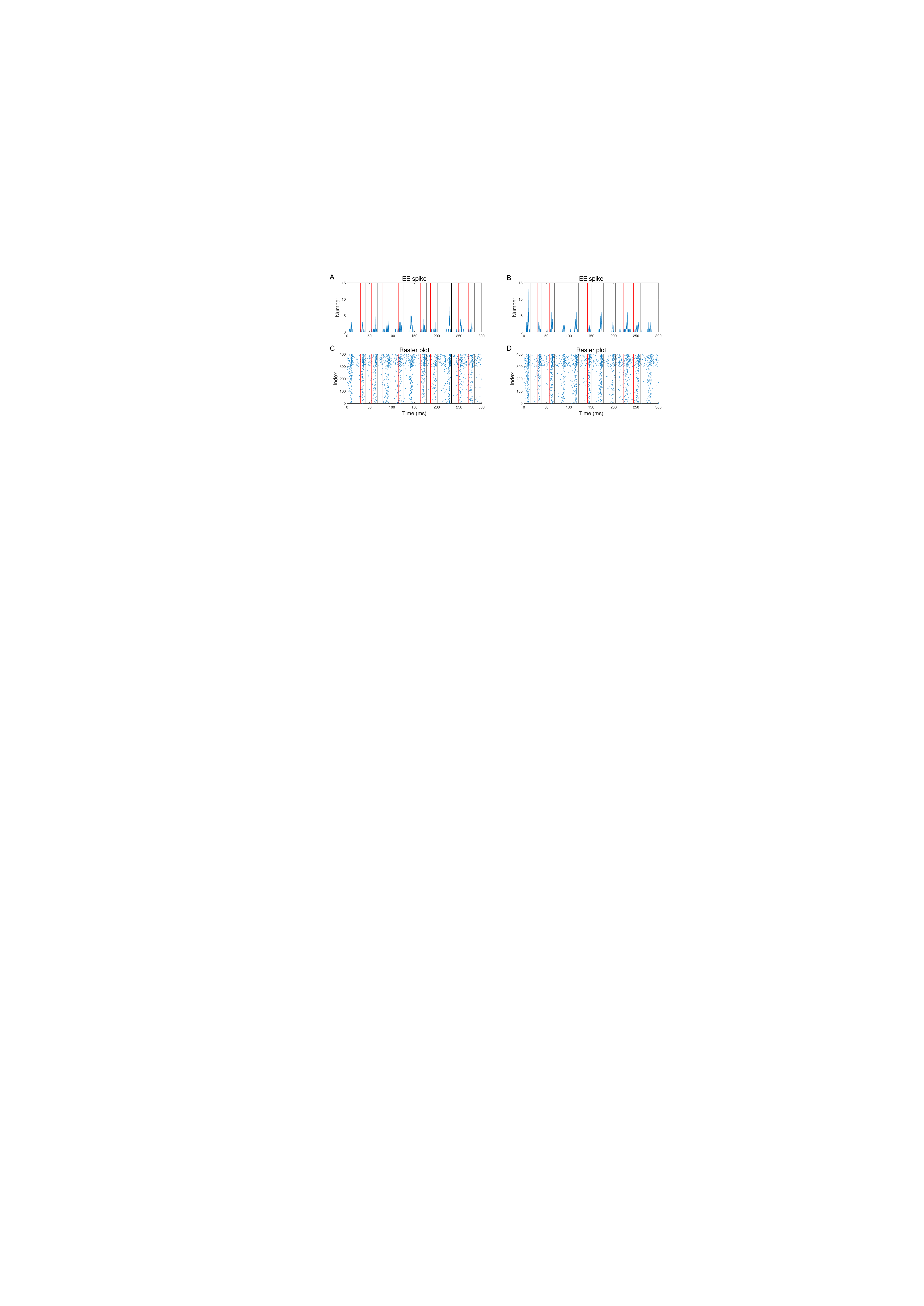}
    \caption{The amount of recurrent excitation serves as the indicator of MFEs. \textbf{A-B.} Number of E spikes caused by recurrent excitation within each 0.25 ms interval. $(S^{EE},S^{IE},S^{EI},S^{II}) = (4, 3, -2.2, -2)$ for \textbf{A}; $(4.44, 2.88, -2.22, -1.70$ for \textbf{B}. \textbf{C-D.} Raster plot of the two simulations. Vertical red and black lines indicate the beginnings and endings of MFEs respectively.}
    \label{Fig2}
\end{figure*}

\subsection{Capturing MFEs from network dynamics}
\label{Sect3.2-CapturingMFE}
The first step of investigating MFE is the accurate detection of MFEs patterns from the temporal dynamics of SNN. 
Due to the lack of a rigorous definition in previous studies, we develop an algorithm to capture MFEs based on the indication of recurrent excitation (Algorithm 1), splitting the spiking patterns into consecutive phases of MFEs and IMIs. 

\heading{The algorithm capturing MFEs.} According to Sect. \ref{Sect3.1-MFEProperties}, the existence of the cascade of recurrent excitation has a causal relation with MFEs. Therefore, given a temporal spiking pattern generated by the SNN, Algorithm 1 detects the initiation of a candidate MFE with the following criteria:
\begin{itemize}
    \item As the influence of an E-to-E excitation, the second E-to-E spike is triggered;
    \item The second E-to-E spike should occur within a 4 ms interval following the first one (twice the excitatory synaptic timescale $\tau_E=2$ms).
\end{itemize}
After the cascade of spiking events, the candidate MFE is deemed terminated if no additional E spike takes place within a 4 ms time-window.
Furthermore, to exclude isolated firing events clustering by chance, we apply merging and filtering processes to each of the MFE candidates.
More specifically, consecutive MFE candidates are merged into one if they occur within 2 ms. Candidates are eliminated if their duration is less than 5 ms {\it or} the number of spikes involved is less than 5. We comment that the filtering threshold is chosen based on the size of 400-neuron networks. A different filtering standard is employed for larger networks, presented below. 

Algorithm 1, based on the timing of recurrent E-spikes, is employed throughout this manuscript to detect MFEs. Two examples, for different choices of parameters, are illustrated in Fig.~\ref{Fig2}, where the initiation and termination of MFEs are indicated by red and black vertical lines. 
In the rest of the paper, we denote the time sections of initiations and terminations of the $m$-th MFE as $t_m$ and $t'_m$. Therefore, the MFE interval is $[t_m, t'_m]$, whereas the IMI after that is $[t'_m, t_{m+1}]$.

\begin{algorithm}[]
\label{MFE-Algo}
\caption{Capturing MFE}
\begin{algorithmic}[5]
\Procedure{Recording, Merging and Filtering MFE }{}
\Repeat
\State A new event in simulation happens
\If {State $==$ Recording MFE}
    \If {$\#$ of EE spikes within previous 4 ms $<$ 2 }
        \State State $\gets$ Not in MFE
        \State Record post-MFE network state $\omega$
    \EndIf
\Else
\If{Two EE spikes occur within 4 ms}
    \State Record at first EE spikes as pre-MFE network state $\omega$
    \EndIf
\EndIf
\Until{current time > terminate time}

\State Merge any consecutive MFE candidates with space $<$ 2 ms 
\For{$i=1$ to number of merged candidates}
\If{(Length of candidate $i$) > 5 ms $\&\&$ (Number of spike) > threshold $\&\&$ (peak $H^{EE}$) > (100 + start $H^{EE}$)}
\State Register candidate $i$ as MFE, output all recorded network state
\EndIf
\EndFor
\EndProcedure
\end{algorithmic}
\end{algorithm}

\section{Learning MFEs with artificial neural networks}
Viewed within the framework of random dynamical systems, SNN dynamics exhibiting $\gamma$-oscillation can be effectively represented by a Poincare map theory \cite{wu2022multi}. 
Consider a solution map of the SNN dynamics depicted in Sect.~\ref{Sect2: MIFModel}, 
\begin{align*}
    \omega_t = \Phi^{\theta}_t(\omega_0, \xi),
\end{align*}
where $\omega_0\in\mathbf{\Omega}$ is the initial condition of the SNN, $\theta\in\mathbf{\Theta}$ indicates SNN parameters, and $\xi$ is the realization of all external/internal noises during the simulation. (The solution map satisfies $\Phi_{t+s} = \Phi_t\circ\Phi_s$.) To study the recurrence of MFEs, we focus on the time sections $\{t_m\}$ on which $\omega_{t_m}$ returns to the initiation of the $m$th MFE. Therefore, we define a sudo-Poincare map:
\begin{equation}
    \omega_{t_{m+1}} = F^{\theta}(\omega_{t_{m}}) = \Phi^\theta_{t_{m+1}-t_m}(\omega_{t_m}, \xi).
\end{equation}
We note that $F^{\theta}$ is not a Poincare map in the rigorous sense, since the initiation of MFEs depends on the temporal dynamics in a short interval (Algorithm \ref{MFE-Algo}). 

The central goal of this paper is to investigate $F^{\theta}$. It is generally impractical to study $F^{\theta}$ in an analytical manner due to the high dimensionality of $\mathbf{\Omega}$. To overcome the difficulty, we propose a \textit{first-principle-based} deep learning framework. 

\subsection{Dissecting $F^{\theta}$ into slow/fast phases}
The first step is to rephrase our goal guided by SNN dynamics. In the regime of $\gamma$-oscillations, the SNN dynamics is composed of the regular alternation of slow/fast phases, though the duration and detailed dynamics of each phases may vary. Therefore, the pseudo-Poincare map $F^{\theta}$ is equivalent to the composition of two maps:
\begin{align}
\label{eqn-4.1-operator-decomp}
    F^{\theta} = F^{\theta}_1 \circ F^{\theta}_2, \quad F^{\theta}_{1,2}: \mathbf{\Omega}\rightarrow \mathbf{\Omega}.
\end{align}
More specifically, $F^{\theta}_1$ maps the network state at the initiation of an MFE ($\omega_{t_m}$) to its termination ($\omega_{t'_m}$), while $F^{\theta}_2$ maps the network state from the beginning of an IMI ($\omega_{t'_m}$) to the initiation of the next MFE ($\omega_{t_{m+1}}$).
We denote $F^{\theta}_{1,2}$ as MFE and IMI mappings, respectively. To summarize, the dynamical flow of $\Phi^t$ is equivalent to:
$$
    ...\,\omega_{t_m}\xrightarrow[\text{MFE}]{F^{\theta}_1} \omega_{t'_m} \xrightarrow[\text{IMI}]{\ ~~~~F^{\theta}_2~~~~\ }  \omega_{t_{m+1}}\xrightarrow[\text{MFE}]{F^{\theta}_1} \omega_{t'_{m+1}}\,...
$$

Our previous studies demonstrated that the slow and relatively quiescent dynamics during IMIs can be well captured by classical coarse-graining methods, i.e., the IMI mapping $F^{\theta}_2$ may be represented by the evolution of certain Fokker-Planck equations \cite{cai2006kinetic}. However, this is not the case for the MFE mapping $F^{\theta}_1$ due to the highly transient and nonlinear dynamics.
Instead, we turn to deep neural networks (DNN) due to its success in tackling high-dimensional and nonlinear problems. In the rest of this section, our goal is to train a DNN with relevant data and generate surrogates of $F^{\theta}_1$. Our DNN has a feedforward structure with 4 layers, consisting of 512, 512, 512 and 128 neurons, respectively.

\subsection{First-principle-based reductions of the problem}
\label{Sect4.2-DimReduction}
Instead of requiring the DNN to learn the full map $F^{\theta}_1$ and directly link the network state from $\omega_{t_m}$ to $\omega_{t'_m}$, we prepare a training set $\mathcal{T}^\theta_{\rm{train}}$ with first-principle-based dimensional reduction, noise elimination, and enlargement for robustness to facilitate the training process.

\subsubsection{Coarse-graining SNN states}
Approximating the features of MFE mapping $F^{\theta}_1$ with a DNN (or any statistical machine learning method) immediately faces the curse of dimensionality. Namely, $F^{\theta}_1$ maps a $3N$-dimensional space $\mathbf{\Omega}$ to itself. To cope with $N$, we propose a coarse-graining model reduction with \textit{a priori} physiological information. 

Instead of enumerating the actual state of every neuron, we assume the E and I neuronal populations form two \textit{ensembles}. That is, the state of a type-Q neuron can be viewed as randomly drawn from a distribution $\bm{\rho}^{Q}(v,H^E, H^I)$ of the $Q$-ensemble, where $Q \in \{E,I\}$. Furthermore, note that there is no fixed architecture of synaptic connection in SNN (see Sect.~\ref{Sect2: MIFModel}) - any neurons in the same population has the same probability to be projected to when a spike occurs. Therefore, it is reasonable to decorrelate $v$ and the $(H^E, H^I)$, i.e.,
\begin{equation}
    \bm{\rho}^{Q}(v,H^E, H^I) \sim \bm{p}^{Q}(v)\cdot\bm{q}^{Q}(H^E, H^I),
\end{equation}
where $\bm{p}^{Q}$ and $\bm{q}^{Q}$ are marginal distributions of $\bm{\rho}^{Q}$.

More specifically, $\bm{p}^{E}(v),\bm{p}^{I}(v)$ yield the distributions of neuronal voltages for both population on a partition of the voltage space $\Gamma$:
\begin{align}
\nonumber
    \Gamma &= \Gamma_1 \cup \Gamma_2 \cup ... \cup \Gamma_{22} \cup \Gamma_{\mathcal{R}}\\
    \label{Eq4.2-Partition}
           &= [-M_r,0) \cup [0,5) \cup ... \cup [M-5,M=100) \cup \{\mathcal{R}\}.
\end{align}
On the other hand, the distribution of pending spike $\bm{q}^{Q}(H^E, H^I)$ is effectively represented by the total number of pending spikes summed over each population, $H^{QE}$ and $H^{QI}$. To summarize, we define a coarse-grained mapping $\mathcal{C}:\bf{\Omega} \mapsto \mathbf{\tilde{\Omega}}$ by projecting the network state $\omega$ onto a 50-dimensional network state $\tilde{\omega}$:
\begin{align}
\nonumber
\mathcal{C}(\omega)=&\,\tilde{\omega} \\
\nonumber
                   = &\left(\bm{p}^{E}(v),\bm{p}^{I}(v), H^{EE}, H^{EI}, H^{IE},
                  H^{II}\right) \\
                  \nonumber
                   =&\left(n^E_1, n^E_2, \cdots, n^E_{22}, n^E_R, n^I_1, n^I_2,  \cdots, n^I_{22}, n^I_R, \right.\\
\label{Eqn-4.2-CGfunction}
                   & \left.H^{EE}, H^{EI}, H^{IE}, H^{II}\right).
\end{align}
Here, $n^Q_k$ denotes the number of type-Q neurons whose potentials lie in $\Gamma_k$, and $\mathbf{\tilde{\Omega}}\subset\mathbb{R}^{50}$ represents the coarse-grained state space. The precise definition of $n^Q_k$ is given in the Appendix.

We now train a DNN to learn the coarse-grained mapping $\widetilde{F}_1^\theta$ with the coarse-grained network states,
$\tilde{\omega}_{t'_m} = \widetilde{F}_1^\theta(\tilde{\omega}_{t_m})$,
where $\tilde{\omega}_{t} = \mathcal{C}(\omega_t)$.
That is, for a set of fixed SNN parameters ($\theta$), the DNN forms a mapping 
$$\widehat{F}^{\theta,\vartheta}_1: \mathbf{\tilde{\Omega}} \mapsto \mathbf{\tilde{\Omega}} $$ 
where $\vartheta$ is the hyperparameter for the trained DNN. From a simulation of SNN dynamics producing $M$ MFEs, each piece of data of the training set is composed by
\begin{itemize}
    \item Input: $x_m = \tilde{\omega}_{t_m}$, the coarse-grained network states at the initiation of the $m$-th MFE, and
    \item Output: $y_m = (\tilde{\omega}_{t'_m},\mathrm{Sp}_E, \mathrm{Sp}_I)$, 
     the coarse-grained state at the $m$-th MFE termination and the numbers of E/I-spikes during MFEs ($m \in \{1,2,..., M\}$).
\end{itemize}
The training process aims for the optimal $\vartheta$ to minimize the following $L^2$ loss function
\begin{align}
\label{Eqn-4.2-lossfunction}
    \mathcal{L}^\theta(\vartheta) = \frac{1}{M} \sum_{m=1}^M \| y_m - \widehat{F}^{\theta,\vartheta}_1 ( x_m ) \|^2_{L^2} \, , 
\end{align}
and we hope to obtain the optimal coarse-grained MFE mapping $\widehat{F}^{\theta}_1$ as an effective surrogate of $\widetilde{F}^{\theta}_1$. 

\subsubsection{Pre-processing: Eliminating high frequency noises}
\label{Sect4.2.2-DCT}

Using a discrete cosine transform (DCT), we pre-process the training data $(x, y)$ by eliminating the noisy dimensions in potential voltage distributions $\bm{p}^{Q}$, i.e., the high frequency modes.  

The raw distributions of membrane potentials $\bm{p}^{Q}(v)$ contain significant high-frequency components (Fig.~\ref{Fig3}A). This is partially due to the stochasticity and the small number of neurons in the MIF model. Unfortunately, the high-frequency components could lead to ``conservative" predictions, i.e., the training of DNN converges to the \textit{averages} of post-MFE voltage distributions to minimize $\mathcal{L}^\theta(\vartheta)$. Therefore, different inputs would yield similar outputs.

To resolve this issue, we apply a DCT method to remove the high frequency components from training data $(x, y)$, only preserving the first eight modes. This is sufficient to discriminate a pair of neurons whose membrane potentials $|v_1-v_2|>5$, where $v_1, v_2 \in \Gamma$. For $|v_1-v_2|<5$, the two neurons are either place in the same or consecutive two bins after coarse-graining. Therefore, the E spikes needed to involve them in the upcoming MFE differs by at most 1. 


More specifically, in the training set, $1.86\times10^5$ pairs of network states $(x, y)$ are collected from a 5000-second simulation of the SNN network (see Eq.~\ref{Eqn-4.2-CGfunction}). For the voltage distribution components of each state, $\bm{p}^{E}(v)$ and $\bm{p}^{I}(v)$, we remove the high-frequency components by
\begin{align}
    \widehat{\bm{p}^{Q}} = \mathcal{F}_c^{-1} \circ T \circ \mathcal{F}_c\{\bm{p}^{Q}\}, \quad Q\in\{E,I\}.
\end{align}
Here, $\mathcal{F}_c$ indicates the DCT operator, and $T$ is a truncation matrix preserving the first eight components. 
Finally, in the training data $(x, y)$, the coarse-grained network state $\tilde{\omega}$ is replaced as
\begin{align*}
\tilde{\omega} = \left(\widehat{\bm{p}^{E}}(v),\widehat{\bm{p}^{I}}(v), H^{EE}, H^{EI}, H^{IE}, H^{II}\right).
\end{align*}
Fig.~\ref{Fig3}AB depicts an example of pre-processing a voltage distribution. We leave the rest of the details regarding the DCT methods to the Appendix.

\subsubsection{Robustness of training}
The robustness of DNN has been addressed in many previous studies \cite{goodfellow2014explaining,yuan2019adversarial}. One approach of improving robustness is via data augmentation, in which the training set is enlarged by adding modified copies of existing data \cite{shorten2019survey}. Motivated by this idea, we propose a training set $\mathcal{T}^\theta_{\rm{train}}$ to account for more irregular inputs than SNN-produced network states, helping the trained DNN generate realistic and robust predictions for MFE dynamics. Based on the pre-processed network states $\tilde{\omega}$, we increase the variability of the first eight frequency modes of $\widehat{\bm{p}^{Q}}(v)$ as they are the most salient information revealed by DCT. 

\begin{figure*}
    \centering
    \includegraphics[width=.9\textwidth]{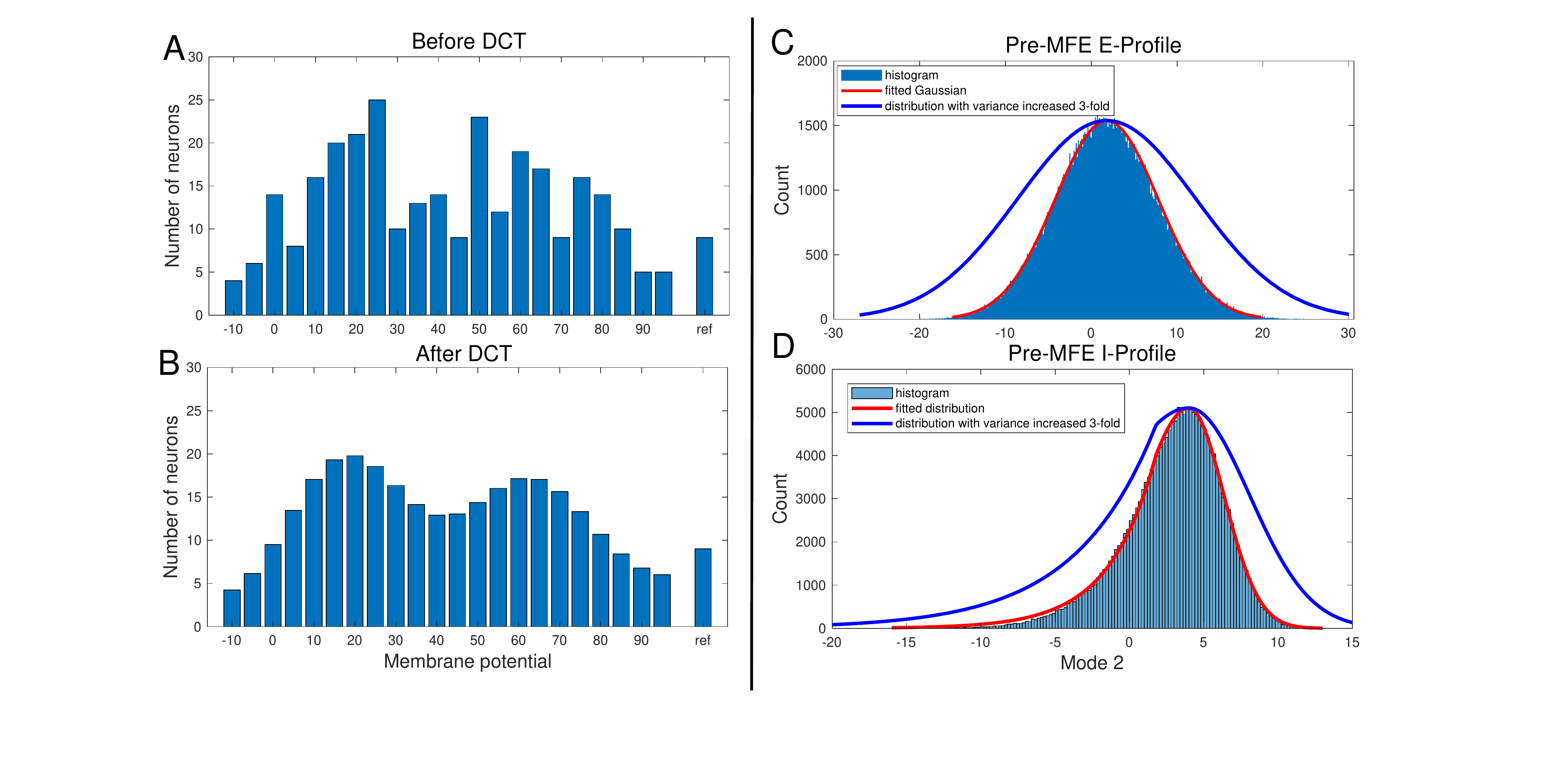}
    \caption{Pre-processing and enlarging the training data-set. \textbf{A.} An example of pre-MFE E-neuron voltage distribution $\bm{p}^{E}(v)$ before DCT + iDCT. For clarity, the number of refractory neurons is depicted by the rightmost bin; \textbf{B.} The pre-processed $\widehat{\bm{p}^{E}}(v)$ after DCT + iDCT; \textbf{C.} The distribution of the 2nd DCT mode of $\widehat{\bm{p}^{E}}(v)$; \textbf{D.} The distribution of the second DCT mode of $\widehat{\bm{p}^{I}}(v)$.}
    \label{Fig3}
\end{figure*}

$\mathcal{T}^\theta_{\rm{train}}$ consists of pairs of pre/post-MFE network states $(x_m, y_m)$. A candidate initial state close to a pre-MFE state $x_m$ in $\mathcal{T}^\theta_{\rm{train}}$ is generated as follows:
\begin{enumerate}
    \item The voltage distributions $\widehat{\bm{p}^{Q}}(v)$. We first collect the empirical distributions of all frequency modes (by concerning each entry of $\mathcal{F}_c\{\bm{p}^{Q}\}$) from $1.86\times10^5$ pre-processed pre-MFE network states. The empirical distributions are fitted by direct combinations of Gaussian and exponential distributions (see, e.g., Fig.~\ref{Fig3}CD). We then increase the variances of fitted distributions by a factor of three, from which the 2nd-8th frequency modes are sampled. On the other hand, the first mode comes from the distributions with the original variances, since it indicates the numbers of neurons outside refractory state. Finally, the higher order of DCT frequency modes are treated as noises and truncated.
    \item The pending spikes $H^{Q'Q}$. Likewise, the ``fit-expand-sample" operations similar to how we treated the 2nd-8th frequency modes in $\widehat{\bm{p}^{Q}}(v)$ are applied to sample the number of pending spikes. 
\end{enumerate}
The factor of three is an arbitrary choice to address the variability of MFE dynamics produced by the SNN, and we leave a more systematic investigation on this issue to future work.

It is also important to make sure all training data are authentic abstractions of SNN states in MFE dynamics, i.e., an MFE can be triggered in a short-time simulation (5ms) 
from a pre-MFE state. 
Therefore, we perform a simulation that begins with
each initial state generated by the enlargement above, and collect the corresponding pre-MFE states ($x$) and post-MFE states ($y$) if an MFE emerges.

In summary, the enlarged $\mathcal{T}^\theta_{\rm{train}}$ consists of $3\times10^5$ pairs of $(x_m,y_m)$:
$$\mathcal{T}^\theta_{\rm{train}} = \left\{(x_m, y_m): m = 1,2,...,3\times10^5\right\},$$
where $50\%$ of the data comes directly from network simulation (after pre-processing of DCT), and the rest comes from the data augmentation process.

\subsection{Generalization for different SNN parameters}
\label{Sect4-networksize}
\heading{Varying the synaptic coupling strengths.} Trained with data from a dynamical system parameterized by $\theta$, DNN produces a surrogate mapping $\widehat{F}^{\theta}_1$. Recall that $\theta$ is a parameter point in a 4D cube of recurrent synaptic coupling strength
\begin{align*}
    \mathbf{\Theta} &= \left\{(S^{EE}, S^{IE}, S^{EI}, S^{II} )\in \right.\\
                & \left.[3.5, 4.5] \times [2.5, 3.5] \times [-2.5, -1.5] \times [-2.5, -1.5]\right\}.
\end{align*}

We treat $\widehat{F}^{\theta}_1$ as a milestone, and further propose a parameter-generic MFE mapping
$$\widehat{F}_1: \mathbf{\tilde{\Omega}} \times \mathbf{\Theta} \mapsto \mathbf{\tilde{\Omega}}. $$ 
That is, given any point $\theta\in\mathbf{\Theta}$ and an pre-MFE state, the DNN predicts the post-MFE state. The optimization problem and loss function are analogous to Eq.~\ref{Eqn-4.2-lossfunction}. 

The training set $\mathcal{T}_{\rm{train}}$ for parameter-generic MFE mapping consists of SNN states from $20000$ different of parameters points, which are randomly drawn from $\mathbf{\Theta}$. 
To ensure reasonable SNN dynamics, each parameter point $\theta$ is tested by the following criteria
\begin{align*}
    &(f_E, f_I) = L(\theta) \\
    &f_E\leq 50 \, \mbox{ Hz and } \,  f_I\leq 100 \mbox{ Hz,} 
\end{align*}
where $L(\theta)$ is a simple linear formula estimating firing rates of E/I populations based on recurrent synaptic weights (for details see Appendix and \cite{li2019well}). For each accepted point in parameter space, we perform a 500 ms simulation of SNN and collect 20 pairs of pre-processed, coarse-grained pre- and post-MFEs states (see Sect.~\ref{Sect4.2-DimReduction}). 


\heading{Varying the network size.} 
Our methods is generic to $\gamma$-oscillations produced by SNNs of different sizes. We demonstrate this on a 4000-neuron SNN (3000 E and 1000 I neurons) sharing all parameters with the previous 400-neuron SNN, except that 
the synaptic weights ($S^{QQ'}$)
are 1/10-th of the values listed in Table~\ref{Table1:Parameters}. This change of synaptic weights aims to control the total recurrent E/I synaptic drives received by each neuron and allows the different network models to have the same mean-field limit. While deferring DNN predictions of SNN dynamics to Sect.~\ref{Sect5-surrogate}, we here note the minor modifications to adopt our methods to the 4000-neuron SNN. 

First, the filtering threshold of MFE is increased from 5 to 50 spikes when collecting MFE-related data (see Sect.~\ref{Sect3.2-CapturingMFE}), since the MFEs are larger. Second, because of the central limit theorem, less intrinsic noises are observed in the SNN dynamics. This leads to an immediate side effect: The span of pre/post-MFE network states collected from SNN simulations is relatively narrower within $\mathbf{\Omega}$, i.e., the training data is less ``general". To compensate, we expand the spans of $\mathcal{T}^\theta_{\rm{train}}$ more courageously. That is, during the enlarging step of training set, the components of pre-MFE states are sampled from distributions with more significant variance expansions.

\subsection{Training result}
The trained DNNs provide faithful surrogate MFE mappings for both $\widehat{F}^{\theta}_1$ and $\widehat{F}_1$.

\begin{figure*}
    \centering
    \includegraphics[width=.89\textwidth]{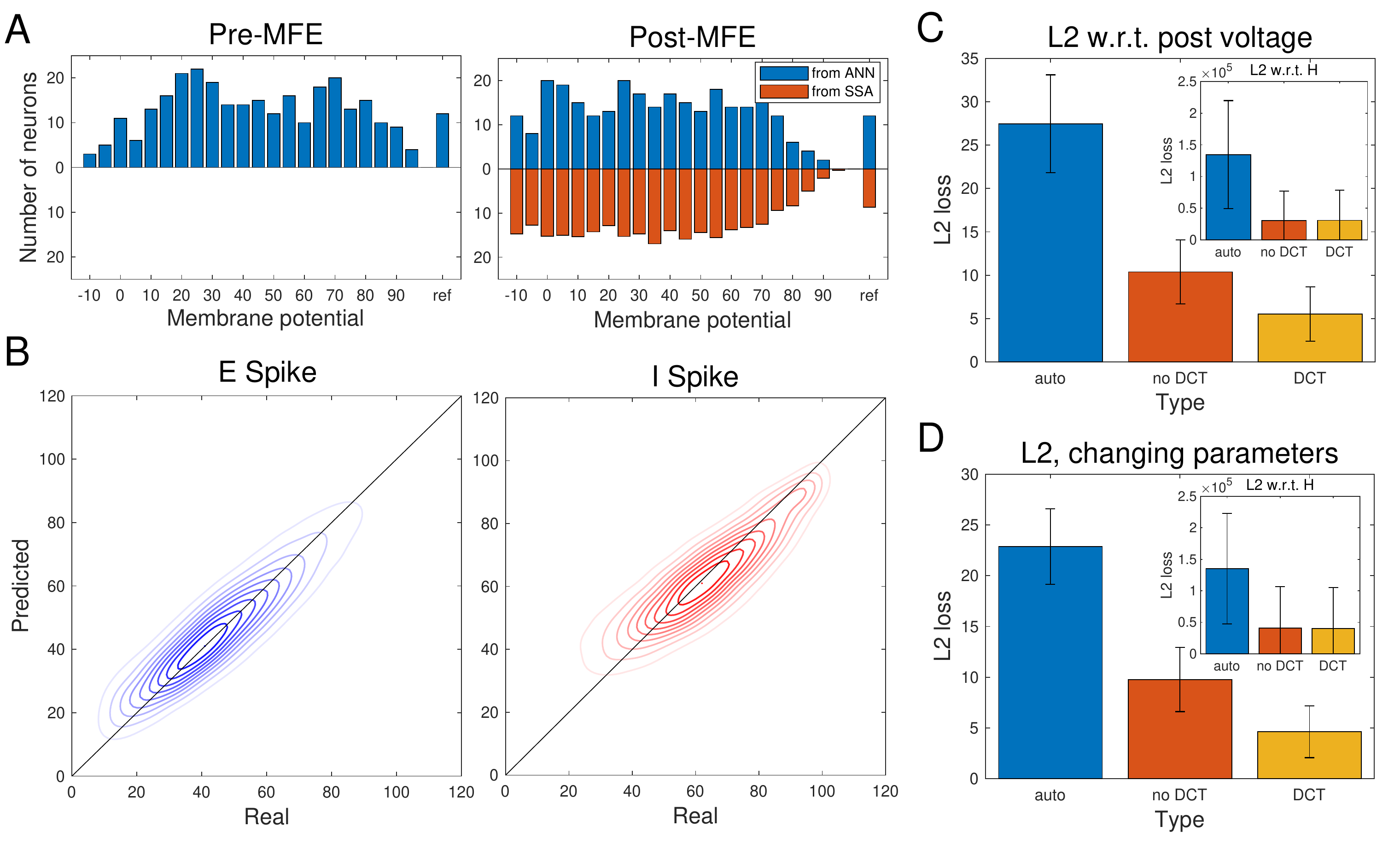}
    \caption{DNNs predictions of post-MFE states. \textbf{A-C:} Mapping $\widehat{F}^{\theta}_1$ for $\theta = (S^{EE}, S^{EI}, S^{IE}, S^{II})= (4, 3, -2.2, -2)$. \textbf{A.} Left: a pre-MFE $\bm{p}^E(v)$; Right: post-MFE $\bm{p}^E(v)$ produced by ANN (blue) vs. spiking network simulations (orange). \textbf{B.} Comparison of E and I spike number during MFEs, ANN predictions vs. SSA simulations. The distributions are depicted by 10-th contours of max in ks-density estimation; \textbf{C.} $L^2$ losses of predicted post-MFE $\bm{p}^Q(v)$ and pending spike number $H$s. Blue: auto-$L^2$ differences between training data; orange: averaged $L^2$-loss without pre-processing; yellow: averaged $L^2$-loss with pre-processing.
    \textbf{D.} $L^2$ loss for the parameter-generic mapping $\widehat{F}_1$. The meaning of all bins are similar to \textbf{B}.}
    \label{Fig4}
\end{figure*}

\heading{Parameter-specific MFE mappings.} 
We first illustrate $\widehat{F}^{\theta}_1$ at a particular parameter point $$\theta = (S^{EE}, S^{EI}, S^{IE}, S^{II}) = (4, 3, -2.2, -2).$$ We test the predictive power of $\widehat{F}^{\theta}_1$ on a testing set
$$\mathcal{T}^{\theta}_{\rm{test}} = \left\{(x'_m, y'_m): m = 1,2,..., M=6\times10^4\right\},$$
where $(x'_m, y'_m)$ are coarse-grained pre/post-MFE states without pre-processing collected from SNN simulations.
In Fig.~\ref{Fig4}A, one example of DNN prediction to post-MFE $\bm{p}^E(v)$ is compared to the SNN simulation results starting from the same $x'_m$. 
Also, the comparison between predicted vs. simulated E/I-spike numbers during MFEs are depicted in Fig.~\ref{Fig4}B. To demonstrate the accuracy of $\widehat{F}^{\theta}_1$,
Fig.~\ref{Fig4}C depicts the $L^2$ losses of different components of post-MFE states. The $L^2$ loss of predicted $\bm{p}^Q(v)$ is $\sim$4, while
the averaged $L^2$ difference between the post-MFE voltage distributions in the testing set is
\begin{align*}
     \underset{1\leq m,\ell\leq M}{\textrm{mean}} \, \| \bm{p}^{E}_m(v) - \bm{p}^{E}_\ell(v) \|^2 + \| \bm{p}^{I}_m(v) - \bm{p}^{I}_\ell(v) \|^2 
    \approx 20\, ,
\end{align*}
Notably, DCT pre-processing effectively improves DNN predictions of voltage distributions by reducing the $L^2$ loss from $\sim$10 to $\sim$4. 
Similar comparison is observed for the prediction of pending spike numbers in the post-MFE states (Fig.~\ref{Fig4}C inset). 

\heading{Parameter-generic MFE mappings.} Likewise, $\widehat{F}_1$ also provides faithful predictions following SNN dynamics in different parameter regimes by labeling training data with synaptic strength parameters. Fig.~\ref{Fig4}D shows similar comparison between $L^2$ losses of different components of post-MFE states.


\section{Producing a surrogate for SNN}
\label{Sect5-surrogate}  
Here, we depict how our ANN predictions provide a surrogate of the spiking network dynamics. We focus on the algorithm to replace $F^\theta$ for a fixed $\theta$. The algorithm for parameter-generic $F$ is analogous. 

Recall that SNN dynamics is divided into a fast phase (MFEs) and a slow phase (IMIs).
Our first-principle-based DNN framework produces a surrogate mapping $\widehat{F}^\theta_1$, and replace the pseudo-Poincare mapping $F^\theta$ if complemented by $F^\theta_2$. We approximate the IMI mapping $F^\theta_2$ by evolving SNN with a \textit{tau-leaping} method, thereby producing the surrogate to the full SNN dynamics by alternating between the two phases.

To resemble the IMI dynamics, we first initialize the network state $\omega$ from the coarse-grained post-MFE state $\tilde{\omega}$ predicted by $F^\theta_1$. Since neurons in the MIF model are exchangeable, $\omega$ can be randomly sampled from $\mathcal{C}^{-1}( \tilde{\omega})$. Specifically, we evenly assign voltage to each type-$Q$ neurons based on $\widehat{p^{Q}}$. On the other hand, same-category pending spikes stay ``pooled" and are assigned to each neuron interchangeably. 
After that, the network state $\omega$ is evolved by a tau-leaping method with 1-ms timesteps. This process is terminated if more than three E-to-E spikes or six E spikes occur within 1 ms, after which the next MFE is deemed to start and the network state $\omega$ is fed to $F^\theta_1$ for another round of prediction\footnote{The different criteria of MFE initiation from Algorithm 1 aims to ensure the robustness of capturing MFE, due to the lack of network state information within each timestep in the tau-leaping simulations.}. The loop $F^\theta$ alternating between $F^\theta_1$ and $F^\theta_2$ is thus closed. The mathematical descriptions of the SNN surrogate algorithm are summarized as follows:
\begin{itemize}
    \item[(i)] $( \tilde{\omega}_{t'_m}, \mathrm{Sp}_E, \mathrm{Sp}_I) = \widehat{F}^\theta_1(x_m)$, where $\tilde{\omega}_{t'_m}$ is the coarse-grained post-MFE state.
    \item[(ii)] Sample $\omega_{t'_m}$ from $\mathcal{C}^{-1}( \tilde{\omega}_{t'_m} )$.
    \item[(iii)] Evolve the network dynamics with tau-leaping method and initial condition $\omega_{t'_m}$. Stop the simulation with terminal value $\omega_{t'_{m+1}}$.
    \item[(iv)] $x_{m+1} = \mathcal{C}(\omega_{t'_{m+1}})$. 
    \item[(v)] Repeat (i)-(iv) with $m = m+1$. 
\end{itemize}

\begin{figure*}
    \centering
    \includegraphics[width=1\textwidth]{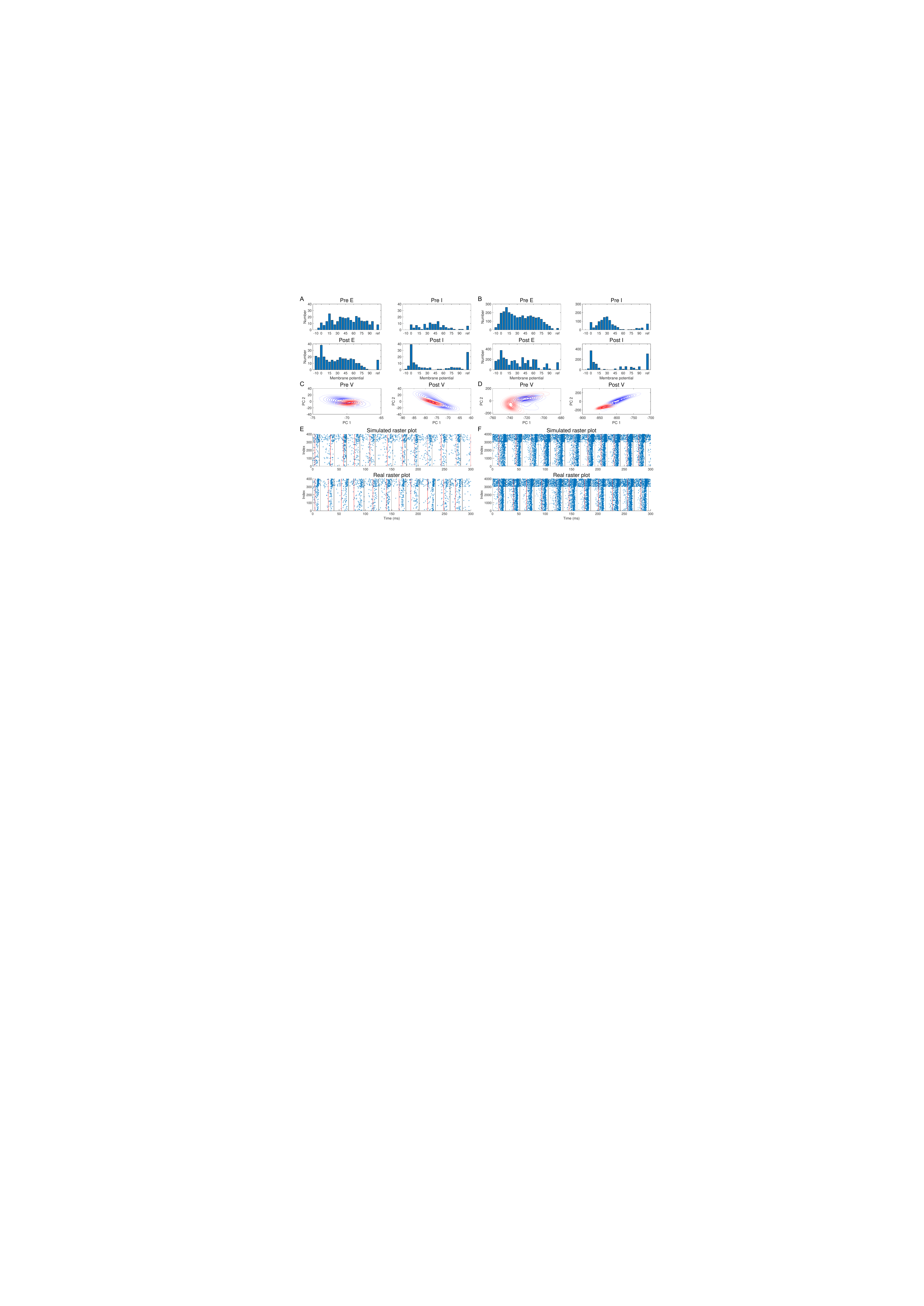}
    \caption{Surrogates of spiking network dynamics produced by the parameter-specific MFE mapping $\widehat{F}^\theta_1$. \textbf{A, C, E:} resembling a 400-neuron spiking network; \textbf{B, D, F:} resembling a 4000-neuron network. \textbf{A, B}: Example of pre and post-MFE voltage distributions $\bm{p}^E$ and $\bm{p}^I$ in the surrogate dynamics. \textbf{C, D}: 10th level curves of the first two principal components of $\bm{p}^E$ and $\bm{p}^I$. (Blue: 20k examples from enlarged training set, red: 2k examples from the surrogate dynamics.) \textbf{E, F}: Raster plots of simulated surrogate dynamics and the real dynamics staring from the same initial profiles. }
    \label{Fig5}
\end{figure*}

\begin{figure*}
    \centering
    \includegraphics[width=1\textwidth]{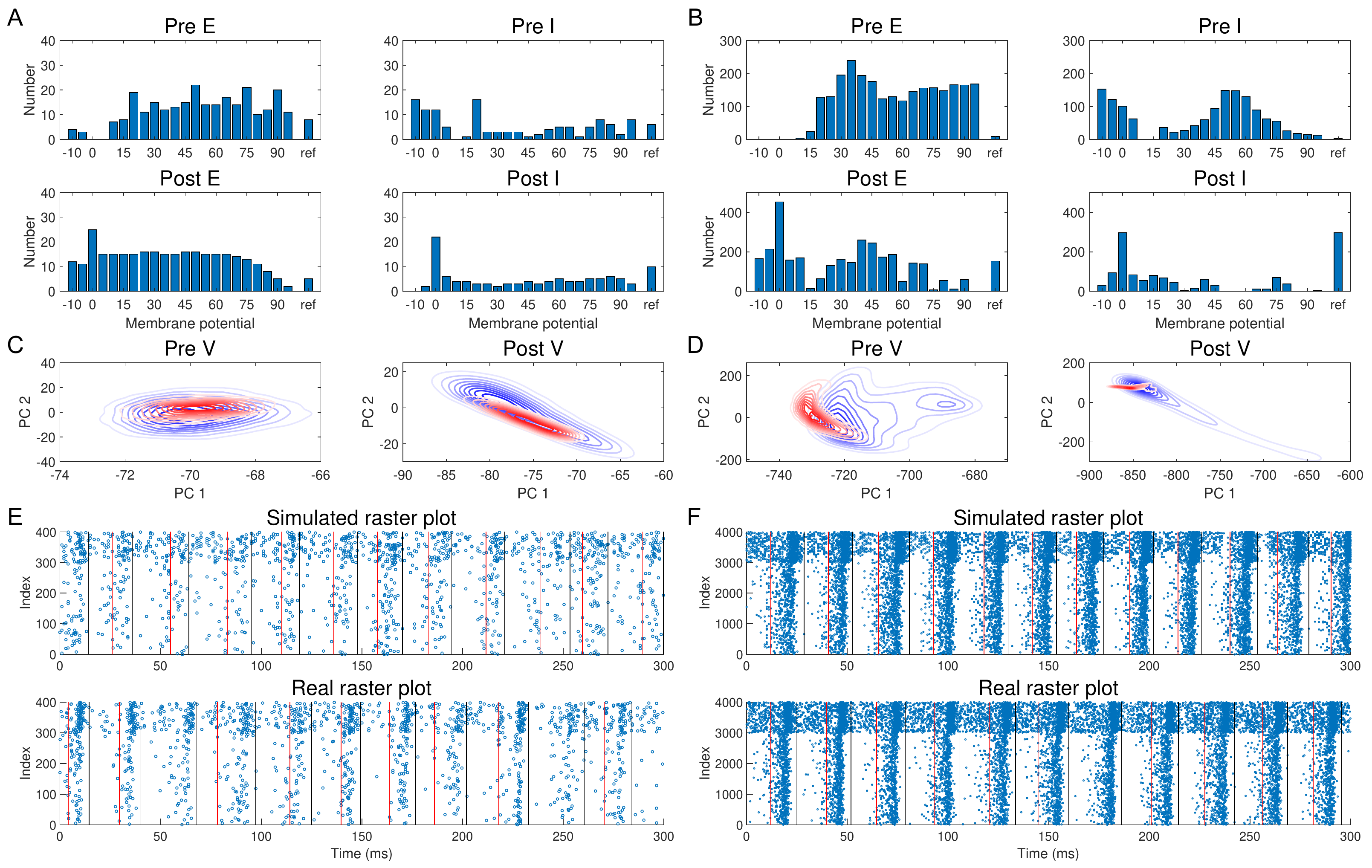}
    \caption{Surrogates of spiking network dynamics produced by the parameter-generic MFE mapping $\widehat{F}_1$. \textbf{A-F} are in parallel to Fig.~\ref{Fig5}.}
    \label{Fig6}
\end{figure*}

We demonstrate the resembling power by producing the raster plot of SNN dynamics labeling all spiking events with neuron index {\it vs.} time in Figs.~\ref{Fig5} and \ref{Fig6}. While the spiking events during IMIs are given by tau-leaping simulation, the spiking patterns during MFEs consists of events uniformly randomly assigned to neurons and times within the MFE interval (with the total number of spikes $(\mathrm{Sp}_E, \mathrm{Sp}_I)$ predicted by our DNN). The durations of the MFEs are randomly sampled from the empirical distribution collected from the SNN simulations. In the rest of the section, we give the details of our surrogate SNN dynamics constructed upon $\widehat{F}^\theta_1$ and $\widehat{F}_1$.

\heading{Parameter-specific predictions.} 
We use the enlarged training set as described in Section 4 to generate $\widehat{F}^\theta_1(x_m)$. Here, $\theta = (S^{EE}, S^{IE}, S^{EI}, S^{II}) = (4, 3, -2.2, -2)$ for the 400-neuron SNN, and the synaptic weights are normalized by 10 times in the 4000-neuron SNN. 

We first focus on the surrogate dynamics of the 400-neuron SNN. Fig.~\ref{Fig5}A gives examples of pre-MFE voltage distributions $\bm{p}^Q$ generated by the tau-leaping simulation and post-MFE $\bm{p}^Q$ generated by DNN predictions. We further compare the distributions of the first two \textit{principal components} of $\bm{p}^Q$ occurring in the surrogate dynamics and the training set (Fig.~\ref{Fig5}B. red: surrogate dynamics; blue: the training set). On the plane of the first two principal components, voltage distributions produced by surrogate dynamics distribute consistently with training sets, suggesting that the surrogate dynamics align very well with the ground-truth network dynamics. The red/blue contours indicate each tenth of the level curves of ks-density of the distributions. We also compare the raster plots of the two dynamics, where the initial SNN conditions are the same (Fig.~\ref{Fig5}E). 

The right half of Fig.~\ref{Fig5} depicts the surrogate dynamics to the 4000-neuron SNN model. The biased distribution of the principal components of $\bm{p}^Q$ is probably due to \textbf{A.} the relatively more narrow training sets (see discussions in Sect. \ref{Sect4-networksize}), and \textbf{B.} the large 1 ms timestep in tau-leaping.

\heading{Parameter-generic predictions.} 
The surrogate dynamics generated by the parameter-generic surrogate MFE mapping $\widehat{F}_1$ and tau-leaping are depicted in Fig.~\ref{Fig6}, whose panels are analogous to Fig.~\ref{Fig5}.  
(The ``real" raster plots of SNN dynamics in Fig.~\ref{Fig6}E is fixed the same as Fig.~\ref{Fig6}E.) The comparable results demonstrate that the DNNs produce faithful surrogate to the SNN dynamics. Interestingly, by comparing panel C and D of Fig.~\ref{Fig5} and \ref{Fig6}, we find that the principal components of $\bm{p}^Q$ in the surrogate dynamics generated by $\widehat{F}_1$ are much less biased than $\widehat{F}^\theta_1$. While a detailed explanation is beyond the current scope of this study, our current conjecture is that the more general training set of $\widehat{F}_1$ helps it deal with the less regular pre-MFE network states in the surrogate dynamics.

\section{Discussions and conclusions}
In this paper, we build an artificial neural network (ANN) surrogate of $\gamma$-dynamics arising from a biological neuronal circuit. The neuronal circuit model is a stochastic integrate-and-fire network that has been well-studied. Similar to many other models \cite{li2019well,zhang2014coarse,rangan2013emergent}, it can exhibit semi-synchronous spiking activities called the multiple-firing events (MFEs), which are transient and highly nonlinear emergent phenomena of spiking neuronal networks. 

In our study, the sensitive \& transient MFE dynamics are represented by the MFE mappings that project the pre-MFE network states to post-MFE network states. The MFE mappings are faithfully approximated by ANNs, despite the significant intrinsic noise in the model. On the other hand, the slower and quieter dynamics between consecutive MFEs are evolved by standard tau-leaping simulations.  
Remarkably, a surrogate of spiking network dynamics is produced by combining ANN approximations and tau-leaping simulations, generating firing patterns consistent with the spiking network dynamics. Furthermore, the ANN surrogate can be generalized to a wide-range of synaptic coupling strengths. 

This paper explores the methodology of learning biological neural circuits with ANNs. In this study, the biggest challenges of developing a successful ANN surrogate are \textbf{A.} processing the high-dimensional, noisy data and \textbf{B.} building a representative training set. Both challenges are addressed by the first-principle-based model reduction techniques, i.e., coarse-graining and discrete cosine transform. The model reductions remove the excessive intrinsic noise. 
The training set collects network states from simulations of SNNs and is enlarged to represent a broader class of voltage distributions. 
Therefore, the training set covers the ``rare" voltage distributions occurring in spiking network dynamics with low probabilities. 

\heading{Future work.} 
The idea of ANN surrogates elaborated in this paper can be extended and applied to other network models. First, many models of brain dynamics share the difficulties of dimensionality, robustness, and generalizability. Therefore, we propose to extend our ideas to model more sophisticated dynamical phenomena of the brain, such as other types of neural oscillations, and to neural circuits with more complicated network architecture. Furthermore, the power of ANNs to handle large data-sets may allow us to extend our framework to deal with experimental data directly. In general, we are motivated by the recent success demonstrating the capability of deep neural networking in representing infinite-dimensional maps \cite{li2020fourier,lu2021learning,wang2021learning}. Therefore, in future work, we suggest exploring more complex network structures (e.g., the DeepONet \cite{lu2021learning}) to build mappings between the states of neural circuits. 

Another interesting but challenging issue is the \textit{interpretability} of ANN surrogates, e.g., relating statistics, dynamical features, and architectures of the spiking network models to the ANNs. A potentially viable approach is to map neurons in the spiking networks to artificial neurons, then examine the connection weight after training. 
However, it is likely that this idea may need ANNs more complicated than the simple feed-forward DNN we considered here. To achieve this goal, one may consider different ANNs with different architectures, such as ResNet or LSTM \cite{he2016deep,hochreiter1997long}. These studies may shed some light on how the dynamics and information flow in neural systems are represented in ANNs. 

\section*{Acknowledgements}
This work was partially supported by supported by the National Science and Technology Innovation 2030 Major Program through grant 2022ZD0204600 (R.Z. Z.W., T.W., L.T.), the Natural Science Foundation of China through grants 31771147 (R.Z., Z.W., T.W., L.T.) and 91232715 (L.T.). Z.X. is supported by the Courant Institute of Mathematical Sciences through Courant Instructorship. 
Y.L. is supported by NSF DMS-1813246 and NSF DMS-2108628.



\section{Appendix}
\subsection{Tau-leaping and SSA algorithms}
In this manuscript, the SNN dynamics are designed to be a Markovian process in phase space $\mathbf{\Omega}$. The simulations are carried out by two algorithms: 
Tau-leaping and Stochastic Simulation Algorithm (SSA). The key difference is that, 
Tau-leaping method processes events that happen during a time step $\tau$ in bulk, while SSA simulates the evolution event by event. Of the two, tau-leaping can be faster (with properly chosen $\tau$), while SSA is usually more precise with the precision that scales with C++ execution. Here we illustrate a Markov jump process as an example.

\heading{Algorithms.} Consider $X(t)=\{x_1(t),x_2(t),..., x_N(t)\}$, where $X(t)$ can take values in a discrete state space 
$$S=\{s_1,s_2,\cdots,s_M\subset\mathbb{R}^N\}.$$ %
The transition from state $X$ to state $s_i$ at time t is denoted as $T_{s_i}^t(X)$, taking an exponential distributed waiting time with rate $\lambda_{s_i\leftarrow X}$. Here, $s_i\in S(X)$ which are states adjacent to state $X$ with a non-zero transition probability. 
For simplicity, we assume $\lambda_{s_i\leftarrow X}$ does not explicitly depend on $t$ except via $X(t)$.  

Tau-leaping only considers $X(t)$ on a time grid $t = jh$, for $j = 0,1,...,T/h$, assuming state transfer occurs for at most one time within within each step:
\begin{align*}
    &P(X^{(j+1)h} = s_i)  \\
    &=\begin{cases}
    h\lambda_{s_i\leftarrow X^{jh}} & \forall s_i\in S(X^{jh}), \\
    1-h\sum_{s_i\in S(X^{jh})}\lambda_{s_i\leftarrow X^{jh}} & s_i = X^{jh}, \\
    0 & \text{otherwise},
    \end{cases}
\end{align*}

On the other hand, SSA accounts for this simulation problem as:
\begin{align*}
  X(T) = T_{X_k}^{t_k}\circ T_{X_{k-1}}^{t_{k-1}}\circ...\circ T_{X_1}^{t_1}(X(0)),
\end{align*}
i.e., starting from $X^0$, $X$ transitions to $X_1, X_2, ..., X_k = X(T)$ at time $0<t_1< t_2< ...< t_k <T $.

For $t_\ell<t<t_{\ell+1}$, we sample the transition time from $\mathrm{Exp}(\sum_{s_i\in S(X(t))} \lambda_{s_i\leftarrow X(t)})$. That is, for independent, exponentially distributed random variables 
$$
\tau_i\sim{\rm Exp}(\lambda_{s_i\leftarrow X(t)}),
$$ 
we have $$
t_{\ell+1} - t_\ell = \min_{s_i\in S(X(t))}\tau_i\sim\mathrm{Exp}(\sum_{s_i\in S(X(t))} \lambda_{s_i\leftarrow X(t)}).$$ 
Therefore, in each step of an SSA simulation, the system state evolves forward by an exponentially distributed random time, whose rate is the sum of rates of all exponential ``clocks". Then we randomly choose the exact state $s_i$ to which transition takes place with probability weighted by the sizes of the pending events. 

\heading{Implementation on spiking networks.} We note that $X(t)$ will changes when 
\begin{enumerate}
    \item neuron $i$ receives external input ($v_i$ goes up for 1, including entering $\mathcal{R}$);
    \item neuron $i$ receives a spike ($H^E_i$ or $H^I_i$ goes up for 1); 
    \item a pending spike takes effect to neuron $i$ ($v_i$ goes up/down according to synaptic strengths);
    \item neuron $i$ walks out from refractory ($v_i$ goes from $\mathcal{R}$ to 0).
\end{enumerate}
The corresponding transition rates are directly given ($\lambda^E$ and $\lambda^I$) or the inverses of the physiological time scales ($\tau^{E}$, $\tau^{I}$, and $\tau^{\mathcal{R}}$). In a SSA simulation, when the state transition elicits a spike in a neuron, the synaptic outputs generated by this spike are immediately added to the pool of corresponding type of effects, and the neuron goes into refractory state. However, in a tau-leaping simulation, the spikes are recorded but the synaptic outputs are processed in bulk at the end of each time step. Therefore, all events within the same time step are uncorrelated.


\subsection{The coarse-graining mapping}
Here we give the definition of the coarse-grained mapping $\mathcal{C}$ in Eq.~\ref{Eqn-4.2-CGfunction}. For 
\begin{align*}
\nonumber
\omega=(&V_1,\cdots,V_{N_E},V_{N_E+1},\cdots,V_{N_E+N_I},\\ 
\nonumber
        &H^E_1,\cdots,H^E_{N_E},H^E_{N_E+1},\cdots,H^E_{N_E+N_I},\\
        &H^I_1,\cdots,H^I_{N_E},H^I_{N_E+1},\cdots,H^I_{N_E+N_I}),
\end{align*}
we define
\begin{align*}
\mathcal{C}(\omega)=\tilde{\omega}=(&n^E_1, n^E_2, \cdots, n^E_{22}, n^E_R, n^I_1, n^I_2, \cdots, n^I_{22}, n^I_R,\\
 &H^{EE}, H^{EI}, H^{IE}, H^{II}),
\end{align*}
where,
\begin{align*}
n^E_i&=\sum_{j=1}^{N_E}\mathbf{1}_{\Gamma_i}(V_j),\quad\text{for }i=1,\cdots,22;\\
n^E_R&=\sum_{j=1}^{N_E}\mathbf{1}_{\{\mathcal{R}\}}(V_j);\\
n^I_i&=\sum_{j=N_E+1}^{N_E+N_I}\mathbf{1}_{\Gamma_i}(V_j),\quad\text{for }i=1,\cdots,22;\\
n^I_R&=\sum_{j=N_E+1}^{N_E+N_I}\mathbf{1}_{\{\mathcal{R}\}}(V_j);
\end{align*}
and
\begin{align*}
H^{EE}&=\sum_{j=1}^{N_E}H_j^E;\qquad H^{IE}=\sum_{j=N_E+1}^{N_E+N_I}H_j^E;\\
H^{EI}&=\sum_{j=1}^{N_E}H_j^I;\qquad H^{II}=\sum_{j=N_E+1}^{N_E+N_I}H_j^I.
\end{align*}
Here, $\mathbf{1}_{\mathrm{A}}(a)$ is an indicator function of set $A$, i.e., $\mathbf{1}_{\mathrm{A}}(a) = 1$ $\forall a\in \mathrm{A}$, otherwise $\mathbf{1}_{\mathrm{A}}(a) = 0$. $\Gamma_i$ is a subset of the state space for membrane potential, and 
$$\Gamma_i = [-15+5i, -10+5i)\cap \Gamma.$$

\subsection{Pre-processing surrogate data: discrete cosine transform}
Here we explain how discrete cosine transform (DCT) works in the pre-processing. For an input probability mass vector 
$$
\bm{p} = (n_1, n_2, \cdots, n_{22}), 
$$ 
its DCT output $\mathcal{F}_c(\bm{p}) = (c_1,c_2,..., c_{22})$ is given by 
\begin{equation}
      c_k = \sqrt{\frac{2}{22}}\sum_{l=1}^{22} \frac{n_l}{\sqrt{1+\delta_{kl}}}\cos\left(\frac{\pi}{2N}(2l-1)(k-1)\right),   
\end{equation}
where $\delta_{kl}$ is the Kronecker delta function. The iDCT mapping $\mathcal{F}^{-1}_c$ is defined as the inverse function of $\mathcal{F}_c$.

\subsection{The linear formula for firing rates}
When preparing the parameter-generic training set, we use simple, linear formulas to estimate the firing rate of E neurons and I neurons ($f_E$ and $f_I$, see \cite{li2019stochastic,li2019well}). We take $\theta \in \mathrm{\Theta}$ for the synaptic coupling strength, while others constants are the same as Table 1.  
\begin{align*}
    &f_E=\frac{\lambda^E(M+C^{II})-\lambda^I C^{EI}}{(M-C^{EE})(M+C^{II})+(C^{EI}C^{IE})}\\
    &f_I=\frac{\lambda^E(M+C^{EE})-\lambda^E C^{IE}}{(M-C^{EE})(M+C^{II})+(C^{EI}C^{IE})}
\end{align*}
where
\begin{align*}
    &C^{EE}=N^EP^{EE}S^{EE},\ 
    C^{IE}=N^EP^{IE}S^{IE}\\
    &C^{EI}=N^IP^{EI}S^{EI},\ 
    C^{II}=N^IP^{II}S^{II}.
\end{align*}

\subsection{Deep network architecture in this paper}
Artificial neural networks (ANNs) are models resembling biological neuronal networks. In general, they are interconnected computation units sending outputs to target units. Many different architectures are possible for ANNs; in this paper, we adopt the \textit{feedforward} deep network architecture, which is one of the simplest (Fig.~8).

\begin{figure}
\label{fig8}
    \centering
    \includegraphics[width=0.4\textwidth]{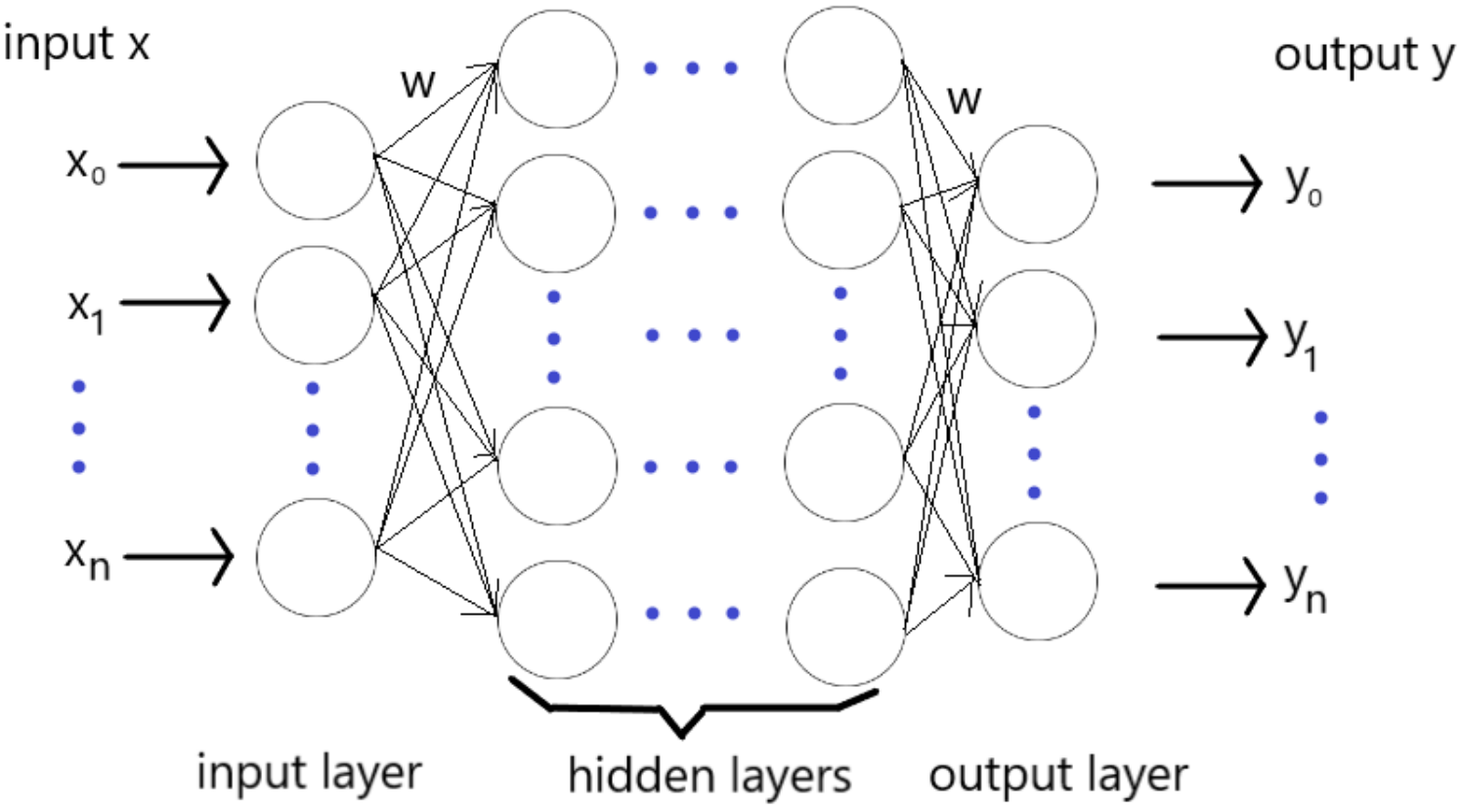}
    \caption{Diagram of a feedforward ANN.}
\end{figure}

A feedforward ANN has a layered structure, where units in the $i-$th layer drives the $(i+1)-$th layer with a weight matrix $\bm{W}_i$ and a bias vector $\vec{b}_i$. Computation is processed from one layer to the next. The "input layer" takes input $\vec{x}$, sending the output $\bm{W}_1x+\vec{b}_1$ to the first "hidden layer"; the first hidden layer then sends output $\bm{W}_2 f_1(\bm{W}_1x+\vec{b}_1)+\vec{b}_2$ to the next layer, and so on, until the "output layer" sends output vector $\vec{y}$. In this paper, we implemented a feedforward ANN with 4 layers, having 512, 512, 512, 128 neurons respectively. We chose the Leaky ReLU function with default negative slope of $0.01$ as our activation function $f_1(\cdot)$.

The training of feedforward ANNs is achieved by the back-propagation (BP) algorithm. Let $\mathcal{NN}(x)$ denote the prediction of the ANN with input $x$, and $L(\cdot)$ the loss function. With each entry $(x,y)$ in the training data, we minimize the loss $l=L(y-\mathcal{NN}(x))$ following the gradients on each dimension of $W_i$ and $b_i$. The computation of gradients takes place from the last layer $W_n$'s and $b_n$, then "propagated back" to adjust previous $W_i$ and $b_i$ on each layer. We chose the mean-square error as our loss function, i.e. $L(\cdot)=||\cdot||_{L^2}^2.$

\begin{figure*}
\label{fig7}
    \centering
    \includegraphics[width=\textwidth]{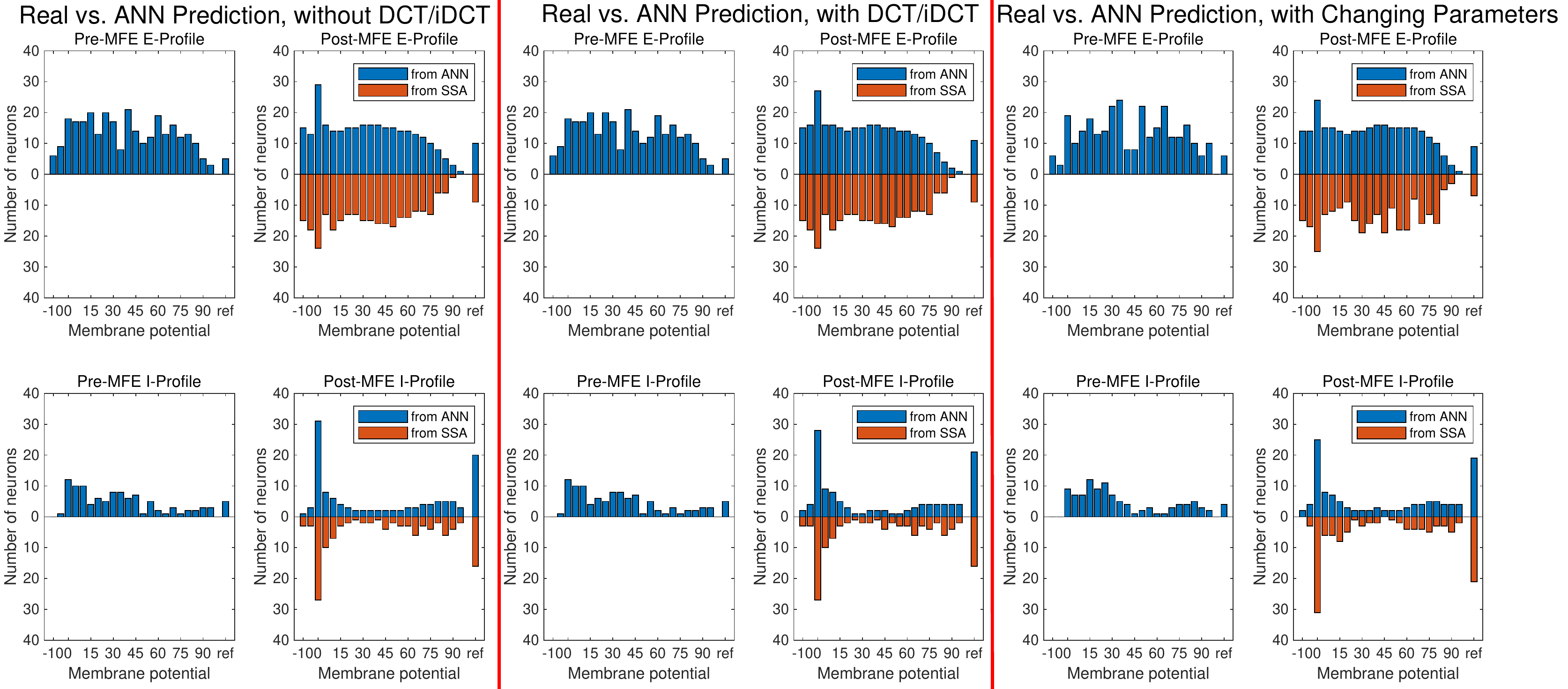}
    \caption{\textbf{Left:} pre-, post- and predicted MFE profiles without DCT + iDCT; \textbf{Middle:} pre-, post- and predicted MFE profiles with DCT + iDCT; \textbf{Right:} pre-, post- and predicted MFE profiles with network parameters as additional inputs of ANN. (Left and Middle: $S^{EE}$, $S^{IE}$, $S^{EI}$, $S^{II}$ = 4, 3, -2.2, -2; Right: 3.82, 3.24, -2.05, -1.87.)}
\end{figure*}
\subsection{Pre-processing in ANN predictions}
As the supplementary information for Fig.~\ref{Fig4}, we compare how ANNs predict post-MFE voltage distributions $\bm{p}^E$ and $\bm{p}^I$ in three different settings in Fig.~8. In each panel divided by red lines, the left column gives an example of pre-MFE voltage distributions, while the right column compares the corresponding post-MFE voltage distributions collected from ANN predictions (red) vs. SSA simulation. Results from ANNs without pre-processing, with pre-processing, and the parameter-generic ANN are depicted in the left, middle, and right panels.

\subsection{Principal components of voltage distributions}
The voltage distribution vectors in the form below are used to plot the distribution in the phase space as shown in middle panels of Fig.~5 and Fig.~6.
\begin{align*}
(&n^E_1, n^E_2, \cdots, n^E_{22}, n^E_R, n^I_1, n^I_2, \cdots, n^I_{22}, n^I_R)   
\end{align*}
The vectors from training set (colored in blue in figures) are selected to generate basis of the phase space through \textbf{svd} function in \textit{numpy.linalg} in python. The first two rows of the $V^\top$ are the first two PCs of the space. The scores of vectors from the training set and approximated results are dot products of these vectors and the normalized PCs. 

The plain ks-density function in Matlab is used to estimate the kernel smoothing density of the profile distribution based on data points generated above. The contours show the level of each tenth of the maximal height (with 0.1\% bias for demonstrating the top) in the distributions.

\onecolumn{
\printbibliography}
\end{document}